\documentclass[12pt]{article}
\usepackage{epsfig}
\usepackage{color}
\usepackage{amssymb,amsmath}
\usepackage{graphicx}
\usepackage{here}
\usepackage{color}
\newcommand{\bea}{\begin{eqnarray}}
\newcommand{\eea}{\end{eqnarray}}

 \makeatletter
\def\alt{\mathrel{\mathpalette\gl@align<}}
\def\agt{\mathrel{\mathpalette\gl@align>}}
\def\gl@align#1#2{\lower.6ex\vbox{\baselineskip\z@skip\lineskip\z@
\ialign{$\m@th#1\hfil##\hfil$\crcr#2\crcr\sim\crcr}}} \makeatother

\begin{document}
\begin{flushright}
\end{flushright}
\vspace*{1.0cm}

\begin{center}
\baselineskip 20pt 
{\Large\bf 
  Neutrino mass square ratio \\
  and neutrinoless double beta decay\\
 in random neutrino mass matrices
}
\vspace{1cm}

{\large 
Naoyuki Haba$^a$, \ Yasuhiro Shimizu$^a$ \ and \ Toshifumi Yamada$^b$
} \vspace{.5cm}

{\baselineskip 20pt \it
$^a$Department of Physics, Osaka Metropolitan University,
Osaka 558-8585, Japan
\\
$^b$Department of Physics, Yokohama National University,
Yokohama 240-8501,Japan
}

\vspace{.5cm}

\vspace{1.5cm} {\bf Abstract} \end{center}

We study the neutrino mass anarchy in the Dirac neutrino, seesaw,
double seesaw models. Assuming the anarchy hypothesis, the mass matrices
are random and distributed in accordance with  the Gaussian measure.
We focus on the distributions of  mass square ratio of the light neutrinos and examine which of these models shows a peak in the probability distribution around the experimental value. We show that the peak position depends on the number of random matrix products. We find that the light neutrino mass hierarchy becomes larger as the number of random matrix products is increased and the seesaw model with the random Dirac and Majorana mass matrices is the most probable to realize the current experimental data.
We also investigate the distributions of the effective Majorana mass for neutrinoless double beta decay. We find that the effective Majorana mass is smaller than the experimental upper bound and tends to be smaller as the number of random matrix products increases because the light neutrino masses become more hierarchical.   We argue that the tendency for lighter neutrino masses to become more hierarchical as the number of products in the random matrix increases can be understood from the probability distribution of singular values in random matrix theory.

\thispagestyle{empty}

\newpage

\setcounter{footnote}{0}
\baselineskip 18pt
%

\section{Introduction}
The Standard Model (SM) of particle physics can mostly explain the experimental results, but there remain several unsolved issues in particle physics. In particular, it was found that the neutrinos have very tiny masses and the flavor mixings are almost maximal in the neutrino oscillation experiments. These properties are quite different from those in the quark sector. Understanding the origin of the neutrino masses has been a major issue.

There have been intensive studies to explain the pattern of the neutrino masses so far. In Ref.\cite{Hall:1999sn}, it was pointed out that the pattern of the  neutrino masses and mixings can be naturally realized if the neutrino mass matrix is random and has no structure. This idea of anarchy is simple and theoretically natural because the coefficients of operators are $O(1)$ if they are allowed by symmetries.  The neutrino mass anarchy has been studied in the literature, see {\it e.g.}, \cite{Haba:2000be,Lu:2014cla, Vissani:2001im, Berger:2000sc, Altarelli:2002sg, deGouvea:2003xe, Agashe:2008fe, Jeong:2012zj, deGouvea:2012ac, Altarelli:2012ia, Brdar:2015jwo,  Ge:2018ofp, Barrie:2019pqc, Jeong:2014qpa, Fortin:2016zyf, Babu:2016aro, Fortin:2017iiw, Long:2017dru, Fortin:2018etr}.
The $O(1)$ parameters are random and treated as statistically distributed. It is important to determine the measure of the $O(1)$ parameters to compare the theoretical predictions with the experimental data. In Ref.\cite{Haba:2000be}, it was pointed out that the neutrino mixings are distributed by the Haar measure because the anarchy requires the basis independence of the neutrino mass matrix. From the Haar measure, it was shown that the large mixings are statistically probable. The eigenvalue distribution of the neutrino matrix is also restricted by the basis independence.
In Ref.\cite{Lu:2014cla}, it was proved that the basis independence uniquely fixes the measure of the random neutrino matrix as the Gaussian distribution. Using the Gaussian distribution, they studied the consequence of the neutrino mass anarchy on cosmology. There are many studies on the neutrino mass anarchy based on the Gaussian distribution \cite{Jeong:2014qpa, Fortin:2016zyf, Babu:2016aro, Fortin:2017iiw, Long:2017dru, Fortin:2018etr}.
  While the idea of anarchy provides a natural realization of the observed neutrino masses and mixings,  it cannot explain the large mass hierarchy and small mixing of quarks and charged leptons. In Ref.\cite{Haba:2000be}, it was shown that a combination of anarchy and approximate $U(1)$ flavor symmetry can explain the hierarchical quark and lepton mass matrices.

In this paper, we extend the anarchy approach to the Dirac neutrino, seesaw \cite{Gell-Mann:1979vob, Yanagida:1979as},  double seesaw models \cite{Mohapatra:1986aw, Mohapatra:1986bd, Ellis:1992zr, Ellis:1992nq, Kang:2006sn}. In these models, the light neutrino mass matrix is given by  the product of different numbers of matrices. If we assume random matrices with Gaussian distribution, the distributions of the light neutrino mixing are almost determined by the Haar measure, that does not depend on the number of random matrix products. However, the distributions of the light neutrino masses depend on the number of random matrix product. Therefore, we focus on the distributions of the neutrino mass square ratio to compare various neutrino models. We perform Monte Carlo analysis to calculate the distributions of the neutrino mass square ratio and examine which of these models shows a peak in the probability distribution around the experimental value.   We show that the peak position depends on the number of random matrix products and  the light neutrino mass becomes more hierarchical  as the number of random matrix products is increased. The distributions of the neutrino mass square ratio were shown for the Dirac neutrino, seesaw models in Ref.\cite{Hall:1999sn,Haba:2000be}, but these focused on the peak positions in each model, and we know that the peak positions depend on the measure of the distribution.  Therefore, it is more important to calculate the probability  that the neutrino mass square ratio agrees  with the experimental data for each model and to compare the probability among the various models.   The results shows that the seesaw model with random Dirac and Majorana masses best explains the current experimental data.
 We also consider neutrinoless double beta decay, which is sensitive to the Majorana nature of neutrinos. We investigate the distributions of the effective Majorana mass  for neutrino double beta decay and show that the effective Majorana mass is below the experimental upper bound  and tends to become smaller as the number of random matrix products increases.    We argue that the tendency for lighter neutrino masses to become more hierarchical as the number of products in the random matrix increases can be understood from the probability distribution of singular values in random matrix theory.

This paper is organized as follows.
In section~2, we explain the random neutrino matrix.
In Section~3, we show the numerical results.
Section~4 summarizes the paper.
\\

\section{Random neutrino mass matrix}

To explain the neutrino masses, we consider the Dirac neutrino model, the seesaw model \cite{Gell-Mann:1979vob, Yanagida:1979as}, and the double seesaw model \cite{Mohapatra:1986aw, Mohapatra:1986bd, Ellis:1992zr, Ellis:1992nq, Kang:2006sn}.

In the Dirac neutrino model, the neutrino mass terms are given by
\bea
{\cal L}= -\overline{\nu}_L m_D \nu_R + h.c.,
\eea
where ${\nu}_L({\nu}_R)$ are the left (right)-handed neutrinos. Here, the observed neutrino mass matrix $m_\nu$ is simply given by
\bea
m_\nu=m_D.
\eea

In the seesaw model, in addition to the Dirac mass term, the Majorana mass term is introduced for the right-handed neutrinos.
\bea
{\cal L}= -\overline{\nu}_L m_D \nu_R - \frac{1}{2}\overline{\nu_R^C} M_R \nu_R +h.c.,
\eea
where $M_R$ is the Majorana neutrino mass matrix.
If $M_D \gg M_R$, the observed light neutrino mass matrix is
given by the seesaw mechanism as follows.
\bea
m_\nu = m_D M_R^{-1} m_D^T. 
\eea

In the double seesaw model, the neutrino mass terms are given by
\bea
{\cal L}= -\overline{\nu}_L m_D \nu_R - \frac{1}{2}\overline{\nu_R^C} M_R \nu_R +-\overline{\nu}_R M S - \frac{1}{2}\overline{S^C} \mu S +h.c.
\eea
where $S$ are extra singlet neutrinos, $M$ and $\mu$ are their Dirac and Majorana mass matrices, respectively. If we assume that there is a hierarchy $M_R >M \gg m_D, \mu$, the light neutrino mass matrix is given by the double seesaw mechanism as follows.

\bea
m_\nu = m_D M^{-1} \mu M^{-1^T} m_D^T. 
\eea

The light neutrino mass can be diagonalized as
\bea
m_\nu =
\begin{cases}
        U m_\nu^{(d)} V ~~~(\mathrm{Dirac}),
\\
        U m_\nu^{(d)} U^T ~~~~(\mathrm{seesaws}),
\end{cases}
\eea
where $U$, $V$ are  unitary matrices and $U$ corresponds to the PMNS matrix in the
charged-lepton diagonal basis. $m_\nu^{(d)}$ is a real diagonal matrix.
\bea
m_\nu^{(d)} = \mathrm{diag}(m_1, m_2, m_3).
\eea
Here, we assume that the order of the eigenvalues is $m_1>m_2>m_3$.
The  light neutrino mass matrix has three mass square differences, $m_1^2-m_2^2$, $m_2^2-m_3^2$,
$m_1^2-m_3^2$, which are relevant to the solar and atmospheric neutrino
oscillations. There are
several definitions for the ratio of two mass square differences.
In Ref.\cite{Haba:2000be}, the ratio is defined as 
\bea
R_1=\begin{cases}
        {\frac{m_2^2-m_3^2}{m_1^2-m_2^2} ~~~ ({m_2^2-m_3^2} < {m_1^2-m_2^2})}.\\
\\
        {\frac{m_1^2-m_2^2}{m_2^2-m_3^2} ~~~ ({m_2^2-m_3^2} > {m_1^2-m_2^2})}.
\end{cases}
\label{R1}
\eea
We also consider another ratio defined by
\bea
R_2=\frac{m_2^2-m_3^2}{m_1^2-m_3^2}.
\label{R2}
\eea
Here, both the definitions satisfy $R_i <1$ and we compare these ratios with the observed neutrino mass square ratio \cite{ParticleDataGroup:2020ssz}.
\bea
\label{Rexp}
R=\frac{\Delta m^2_\mathrm {solar}}{\Delta m^2_\mathrm{ atm}}=(2.96\pm 0.08) \times 10^{-2}.
\eea

If the light neutrino masses have  Majorana nature, the neutrinoless double-beta decay can occur and the decay amplitude is proportional to the effective Majorana neutrino mass defined by

\bea
\label{mee}
m_{ee}=\left| \sum_i (U_{PMNS})_{ei}^2 \, m_i \right|.
\eea
The neutrinoless double beta decay has been searched using various nuclei.  The best upper bound on $m_{ee}$ has been obtained from the KamLAND-Zen experiment as follows \cite{KamLAND-Zen:2016pfg}.
\bea
\label{KamLAND}
m_{ee} < (0.061-0.165)~\mathrm{eV}.
\eea
The future experimental sensitivity is expected $m_{ee}\simeq 0.01$ eV \cite{Agostini:2022zub}.

To study the probability distributions of physical quantities for neutrinos, we should determine the measure of the mass matrices. The anarchy hypothesis requires that the neutrino mass matrix is basis independent. In Ref.\cite{Haba:2000be}, it was shown that the measure of the neutrino mixing is uniquely determined by the Haar measure and that the large neutrino mixings are highly probable.
In addition to the measure of the mixing, the basis independence restricts the measure of the eigenvalues. In Ref.\cite{Lu:2014cla}, it was proven that the basis independence requires that the measure of the neutrino matrix should obey the Gaussian distribution.
For the Dirac mass matrix, which is a general complex matrix, the measure is given by
\bea
\label{eq:Dirac}
dm_D&=&\prod_{ij} dm_{D,ij} ~e^{-\mathrm{ tr}(m_D m_D^\dagger )} . 
\eea
For the Majorana mass matrix, which is a symmetric complex matrix, the measure is
given by
\bea
\label{eq:Majorana}
dM_R&=&\prod_{i\ge j} dM_{R,ij} ~e^{-\mathrm{ tr}(M_R M_R^\dagger )} .
\eea
Notice that the basis independence does not restrict the overall scale factor and these measures are defined up to an overall scale factor of the mass matrix.
This scale factor is used to fix the neutrino mass scale as in Ref.\cite{Lu:2014cla}.

\section{Numerical Results}
We investigate the probability distributions for the following five cases of the light neutrino mass. 

\begin{enumerate}
\item Dirac-type neutrino mass.
\bea
m_\nu = m_D.
\label{Dirac}
\eea
The measure of $m_D$ is Gaussian in Eq.(\ref{eq:Dirac}). For the numerical calculation, we need to fix the overall scale of random matrices. However, the mass square ratios do not depend on the choice of the overall scale factor. We chose the overall scale of $m_D$ to fix the mass square difference at $\Delta m^2_\mathrm {atm}=2.5 \times 10^{-3}$. 

\item Seesaw model with a diagonal right-handed neutrino mass matrix.\\
  To explain the large mixing in the neutrino sector, the anarchy
  in the Dirac mass matrix is enough. The right-handed neutrinos are gauge singlets, and they may have different flavor structure.
  For simplicity, we consider the diagonal form, $M_R=M_R^{(0)}{\bf 1}$ and 
  the light neutrino mass matrix is given by
\bea
  m_\nu = \frac{1}{M_R^{(0)}} m_D m_D^T. 
\eea
The measure of $m_D$ is assumed Gaussian in Eq.(\ref{eq:Dirac}). We choose
$M_R^{(0)}=10^{12}$ GeV and the overall scale of $m_D$ to fix the mass square difference at $\Delta m^2_\mathrm{atm}$.

\item Seesaw model. \\
\bea
m_\nu = m_D M_R^{-1} m_D^T .
\label{Seesaw}
\eea
  
$m_D$ and $M_R$ are assumed random matrices and the measures are Gaussian in Eq.(\ref{eq:Dirac}), (\ref{eq:Majorana}), respectively. We choose the overall the overall scale of $M_R$ to be $10^{12}$ GeV and that of $m_D$ to fix the mass square difference at $\Delta m^2_\mathrm{atm}$.

\item Double Seesaw model with a diagonal singlet neutrino mass matrix.\\
    For simplicity, we assume that $\mu=\mu^{(0)}{\bf 1}$ and the light neutrino mass matrix is given by
\bea
m_\nu = \mu^{(0)} m_D M^{-1}M^{-1T}m_D^T.
\label{DoubleSeesaw}
\eea
The measures of $m_D$ and $M$ are Gaussian in Eq.(\ref{eq:Dirac}), (\ref{eq:Majorana}), respectively. We choose $\mu^{(0)}=0.2$ GeV and the overall scale of $M$ to be 30 GeV and that of $m_D$ to fix the mass square difference at $\Delta m^2_\mathrm{atm}$.

\item Double Seesaw model.
\bea
m_\nu = m_D M^{-1}\mu M^{-1T}m_D^T.
\label{DoubleSeesaw2}
\eea
Here the measures of $m_D$ and $M$ are Gaussian in Eq.(\ref{eq:Dirac})
and the measure of $\mu$ obeys Eq.(\ref{eq:Majorana}). We choose the overall scale of $\mu$ $(M)$ to be $0.2$ $(30)$ GeV and that of $m_D$ to fix the mass square difference at $\Delta m^2_\mathrm{atm}$.
\end{enumerate}
Notice that these five cases correspond that the light neutrino mass is
composed of the product of one, two, three, four, five random matrices, respectively.

We generated $10^6$ data sets assuming the Gaussian distributions for a random mass matrix for each case. Then, we calculated the light neutrino mass matrix and investigate and the probability distributions of the ratios $R_i$.
There are several literatures where the distributions of the neutrino mass square ratio \cite{Hall:1999sn,Haba:2000be} and the effective Majorana mass for the neutrinoless double beta decay \cite{Lu:2014cla}, but these focused on the peak positions in each model. Since the peak positions depend on the measure of the distribution, it is more important to calculate the probabilities that agree  with the experimental data for each model.

\begin{figure}[htbp]
\begin{center}
\includegraphics[width=9cm]{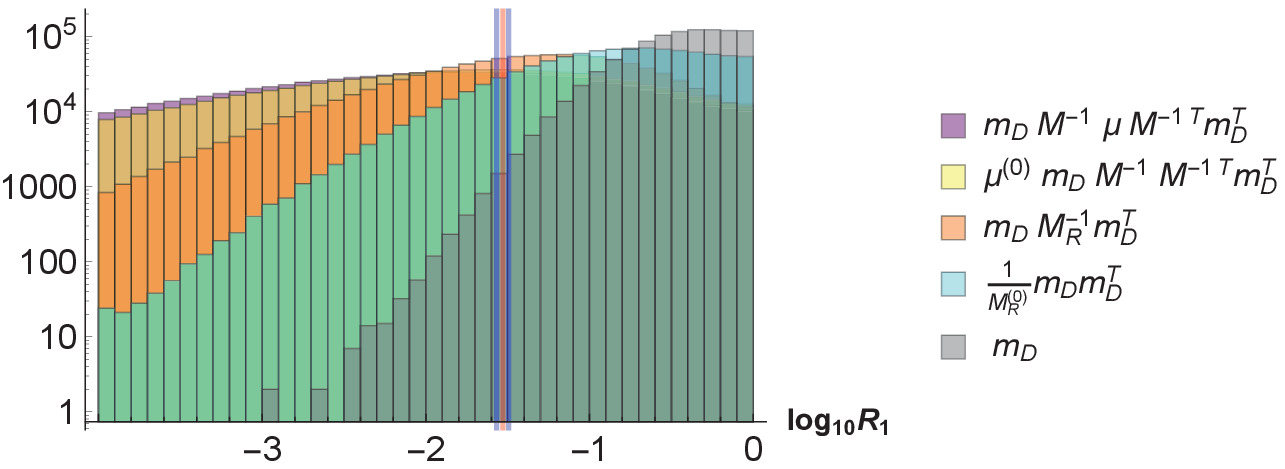}\includegraphics[width=9cm]{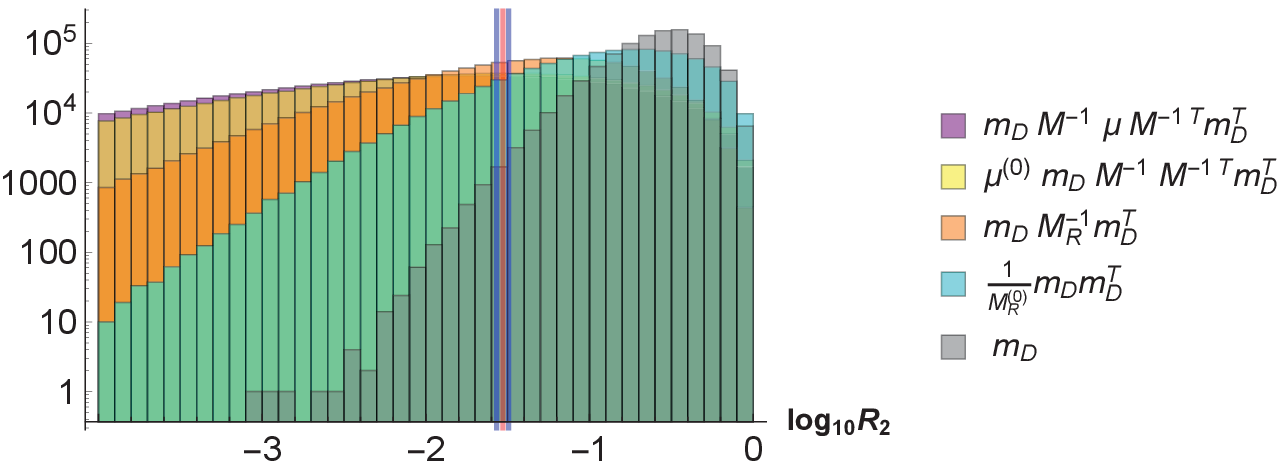}
\caption{Histograms of the ratios $R_1$ (left) and $R_2$ (right) in each model. The red and blue lines correspond to the central value and the 3 $\sigma$ range of the current experimental value, respectively.}
\label{fig1}
\end{center}
\end{figure}

\begin{figure}[htbp]
   \includegraphics[width=18cm]{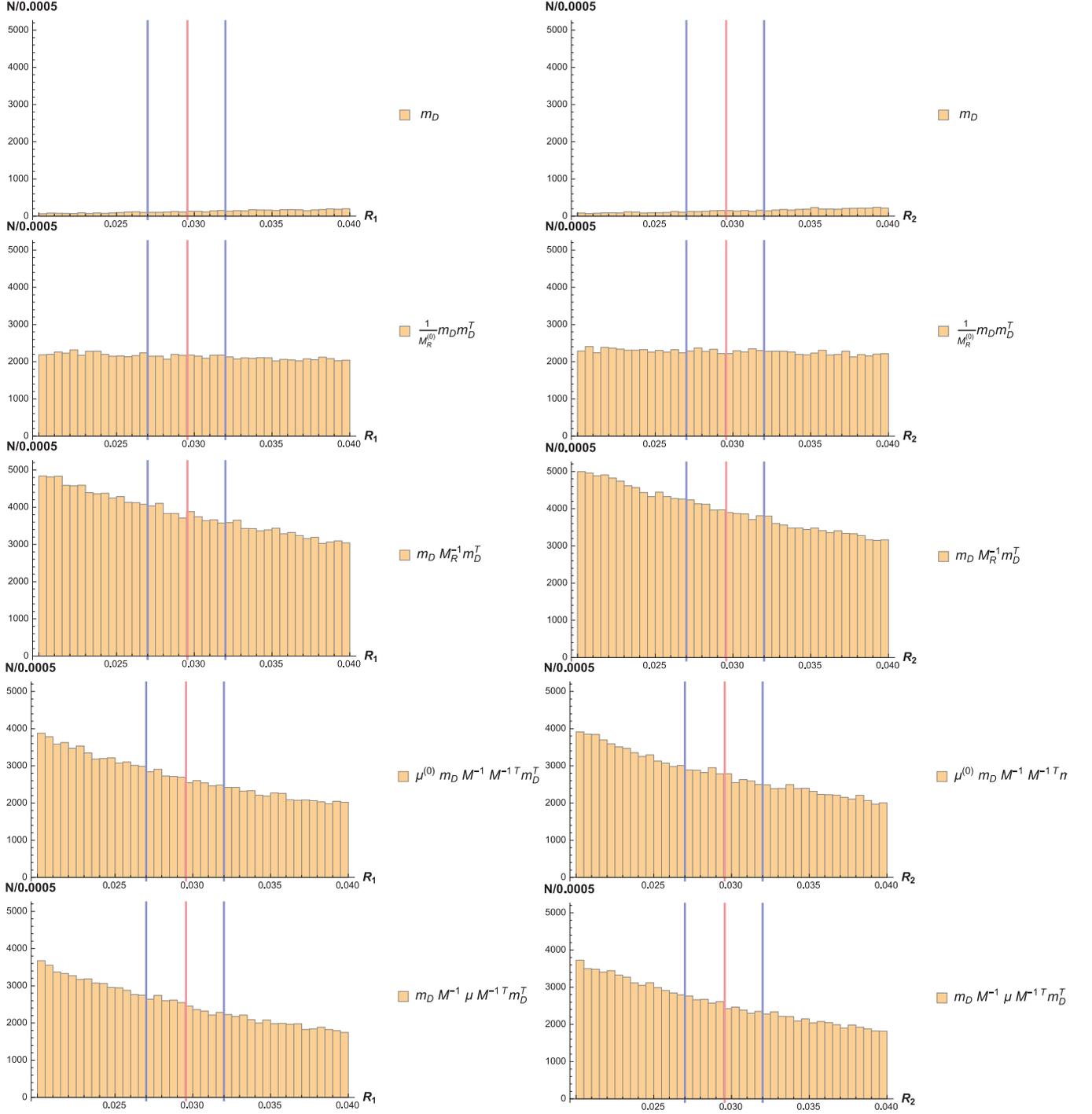}
\begin{flushleft}
\caption{Histograms of the ratios $R_1$ (left) and $R_2$ (right) around the current experimental value. The red and blue lines correspond to the central value and the 3 $\sigma$ range of the current experimental value, respectively.
}
\label{fig2}
\end{flushleft}
\end{figure}

Fig.\ref{fig1} shows the histograms for $R_i$ for the five cases.
Comparing the histograms for $R_1$ and $R_2$, we can see that the shapes are different around $R_i\sim 1$. We find that the position of the peak in each histogram becomes smaller,  {\it i.e.}, the light neutrino masses become more hierarchical, as the number of random matrix products increased for both $R_i$. 
Fig.\ref{fig2} shows the histograms for $R_i$ near the current experimental region in Eq.(\ref{Rexp}). 
For comparison, all vertical and horizontal axes in the histograms are fixed in the same way.
We can see that the histograms around the experimental region have similar shapes for both $R_i$.
Table.\ref{prob} shows the probabilities that $R_i$ falls within the experimental 3 $\sigma$ region. We can see that the probability is almost the same as using either $R_i$. Note that each probability
is not large because the mass square ratio has already been measured rather precisely.
It is important to compare the probabilities among the models and  it turns out that the seesaw model with the random Dirac and Majorana matrices provides the highest probability for both $R_i$. It is interesting that the seesaw model is the most probable from the anarchy approach. 

\begin{table}[H]
\begin{center}
  \caption{Probabilities that $R_i$'s fall within the 3 $\sigma$ range of experimental values in each neutrino mass matrix. Numbers are written in percentages.}
  \begin{tabular}{|c||c|c|c|c|c|} \hline
     & $m_D$ & $(1/M_R^{(0)})m_D m_D^T$ & $m_D M_R^{-1}m_D^T$ & $\mu^{(0)} m_D M^{-1}M^{-1T}m_D^T$ & $m_D M^{-1}\mu M^{-1T}m_D^T$  \\ \hline\hline
    $R_1$ & 0.1 & 2.2 & 3.8  & 2.7  & 2.5  \\\hline
    $R_2$  & 0.1 & 2.3 & 4.0 & 2.7  & 2.5  \\ 
     \hline
    \end{tabular}
  \label{prob} 
  \end{center}
\end{table}

\begin{figure}[htbp]
\begin{center}
\includegraphics[width=9cm]{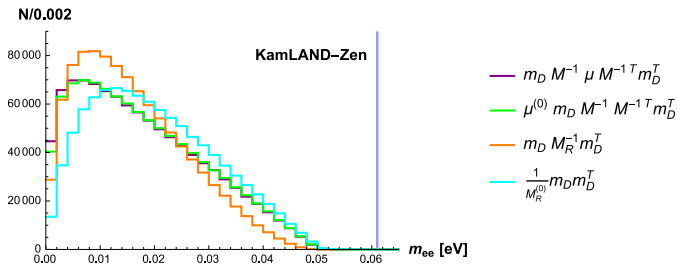}
\caption{Distributions of the effective Majorana mass for the four cases. The blue vertical line corresponds to the best upper bound by the KamLAND-Zen experiment.}
\label{fig3}
\end{center}
\end{figure}

Fig.\ref{fig3} shows the distributions of the effective Majorana neutrino mass for the four cases.\footnote{
  In Ref.\cite{Lu:2014cla}, the distributions of the effective Majorana neutrino mass has been shown for the seesaw model with random Dirac and Majorana mass matrices.
}
Here $m_{ee}=0$ for the Dirac-type neutrino case.
It can be seen that the distributions are below the current experimental bound.
  We find that the position of the peak in each distribution decreases as the number of random matrix products increased, which is  similar to the behavior of the mass square ratio,
because the light neutrino masses become more hierarchical.
  Table.\ref{prob2} shows that the effective Majorana mass is larger than the future experimental reach, $m_{ee}>0.01$ eV.

\begin{table}[H]
\begin{center}
  \caption{Probabilities that the effective Majorana mass is larger than a future experimental reach, $m_{ee}>0.01$ eV. Numbers are written in percentages.}
  \begin{tabular}{|c||c|c|c|c|c|} \hline
     & $m_D$ & $(1/M_R^{(0)})m_D m_D^T$ & $m_D M_R^{-1}m_D^T$ & $\mu^{(0)} m_D M^{-1}M^{-1T}m_D^T$ & $m_D M^{-1}\mu M^{-1T}m_D^T$  \\ \hline\hline
    $m_{ee}$  & 0 & 78 & 67 & 69  & 68  \\ 
     \hline
    \end{tabular}
  \label{prob2} 
  \end{center}
\end{table}

The tendency for the light neutrino masses to become more hierarchical as the number of products in the random matrix increases can be understood from the probability distribution of singular values in random matrix theory. Let us consider the singular values of the product $P_M$ of $M$ independent complex random matrices $X_i$ of size  $N\times N$,
\bea
P_M=X_MX_{M-1}...X_1,
\label{PM}
\eea
which can be written by the singular decomposition, $P_M=V_M\Lambda_M U_M$, where $V_M$ and $U_M$ are unitary matrices and $\Lambda_M=\mbox{diag}(\lambda^{(M)}_1,\ldots,\lambda_N^{(M)})$ contains the singular values. 
The partition function of squared singular values $s_a=(\lambda_a^{(M)})^2$, $a=1,..,N$, is given by analytically \cite{Akemann:2013},
\bea
{\cal Z}_N^{(M)}&=& 
\int_0^\infty \prod_{a=1}^N ds_a {\cal P}^{(M)}_{\text{jpdf}}(s)\ ,
\ \ \ \ \
\label{partition}\\
{\cal P}^{(M)}_{\text{jpdf}}(s)&=& {C}_{N}^{(M)} \Delta_N(s)
\det_{1\leq c,d\leq N}\left[
G^{M,\,0}_{0,\,M}\left(\mbox{}_{0,\ldots,0,d-1}^{-} \bigg| \, s_c\right)\right]\ .
\label{jpdf}
\eea
Here $({C}_{N}^{(M)})^{-1}= N!\prod_{a=1}^{N}\Gamma(a)^{M+1}$ is the normalization constant and
$\Delta_N(s)$ is the Vandermonde determinant, $\Delta_N(s) =\prod_{N\geq a>b\geq1}(s_a-s_b)$, and $G_{p,q}^{m,n}\left(\begin{array}{cccc}                                                                a_1, & a_2, & \ldots, & a_p \\
b_1, & b_2, & \ldots, & b_q
\end{array}\biggl|z
\right)$ is the Meijer $G$-function defined by \cite{Gradshteyn}
\bea
G_{p,q}^{m,n}\left(\begin{array}{cccc}                                                                a_1, & a_2, & \ldots, & a_p \\
b_1, & b_2, & \ldots, & b_q
\end{array}\biggl|z
\right)=\frac{1}{2\pi i}\int\limits_{\cal C}du\,\frac{\prod_{j=1}^m\Gamma(b_j-u)\prod_{j=1}^n\Gamma(1-a_j+u)}
{\prod_{j=m+1}^q\Gamma(1-b_j+u)\prod_{j=n+1}^p\Gamma(a_j-u)}\,z^u\ .
\label{MeijerG}
\eea
Although the light neutrino mass matrices in Eqs.(\ref{Dirac})-(\ref{DoubleSeesaw2}) do not have  the same form as Eq.(\ref{PM}), the singular value distributions of the neutrino mass matrices are expected to have similar singular value distributions to Eq.(\ref{partition}). To compare with the neutrino mass square ratio $R_2$, we consider the ratio of singular values for $N=3$ as follows.
\bea
R_s =\frac{s_3-s_2}{s_3-s_1},
\eea
where $s_1\le s_2\le s_3$. Using the partition function Eq.(\ref{partition}), the mean value of $R _s$ is given by
\bea
\langle R^{(M)}_s \rangle
= \int_0^\infty \prod_{a=1}^{N=3} ds_a {\cal P}^{(M)}_{\text{jpdf}}(s) R_s\ .
\eea
We obtained $\langle R^{(M)}_s \rangle=$ 0.40, 0.19, 0.12 and 0.094 for $M=1, 2, 3$ and 4, respectively. The mean values become larger as $M$ increases, which indicates that the singular values become more hierarchical as the number of random matrix products is increased. This behavior is consistent with the result of the neutrino mass hierarchy.

\section{Summary}

We have studied the neutrino mass anarchy in the Dirac neutrino, seesaw,
double seesaw models. The anarchy approach requires that the mass matrix is basis independent and this leads that the measure of the matrix should obey the Gaussian distribution. We have considered several neutrino models, in which the light neutrino mass matrix is composed of the product of several random matrices.
We have focused on the neutrino mass square ratio and the effective Majorana mass for the neutrinoless double beta decay and examined which of these models shows a peak in the probability distribution near the experimental value.
There are several literatures where the distributions of the neutrino mass square ratio and the effective Majorana mass for the neutrinoless double beta decay, but these focused on the peak positions in each model. Since the peak positions depend on the measure of the distribution, it is more important to calculate the probabilities that agree  with the experimental data for each model.

We have calculated the probability distributions by the Monte-Carlo method, assuming that the measure of the random matrices is Gaussian.
We have found that the distribution of the light neutrino mixing is essentially  independent of the number of random matrix products. However, the eigenvalue distribution of the light neutrino mass depends on the number.
We have presented the probability distributions of the mass-square ratios $R_i$ in each neutrino model. We found that the location of the peak becomes smaller and the light neutrino masses become more hierarchical as the number of random matrix products increases.
We have calculated the probabilities that  $R_i$ falls within the 3 $\sigma$ range of the experimental data and found that the probability becomes maximal for the seesaw model with random Dirac and Majorana mass matrices.
  We also have investigated the distributions of the effective Majorana mass for neutrinoless double beta decay. We have shown that the distributions are below the experimental upper bound
  and found that the effective Majorana mass tends to be smaller as the number of random matrix products increases because the light neutrino masses become more hierarchical.
    It has been argued that the tendency for lighter neutrino masses to become more hierarchical as the number of products in the random matrix increases can be understood from the probability distribution of singular values in random matrix theory.

\section*{Acknowledgement}

This work is partially supported by Scientific Grants by the Ministry of Education, Culture,Sports, Science and Technology of Japan, Nos. 17K05415, 18H04590 and 19H051061 (NH), and No. 19K147101 (TY). The authors would like to thank Professor Tsutomu Yanagida for drawing our attention to the neutrinoless double beta decay.


\begin{thebibliography}{99}
\bibitem{Hall:1999sn}
L.~J.~Hall, H.~Murayama and N.~Weiner,
Phys. Rev. Lett. \textbf{84}, 2572-2575 (2000)
doi:10.1103/PhysRevLett.84.2572
[arXiv:hep-ph/9911341 [hep-ph]].



\bibitem{Haba:2000be}
N.~Haba and H.~Murayama,
Phys. Rev. D \textbf{63}, 053010 (2001)
doi:10.1103/PhysRevD.63.053010
[arXiv:hep-ph/0009174 [hep-ph]].

\bibitem{Lu:2014cla}
X.~Lu and H.~Murayama,
JHEP \textbf{08}, 101 (2014)
doi:10.1007/JHEP08(2014)101
[arXiv:1405.0547 [hep-ph]].


\bibitem{Vissani:2001im}
F.~Vissani,
Phys. Lett. B \textbf{508}, 79-84 (2001)
doi:10.1016/S0370-2693(01)00485-3
[arXiv:hep-ph/0102236 [hep-ph]].

\bibitem{Berger:2000sc}
M.~S.~Berger and K.~Siyeon,
Phys. Rev. D \textbf{63}, 057302 (2001)
doi:10.1103/PhysRevD.63.057302
[arXiv:hep-ph/0010245 [hep-ph]].

\bibitem{Altarelli:2002sg}
G.~Altarelli, F.~Feruglio and I.~Masina,
JHEP \textbf{01}, 035 (2003)
doi:10.1088/1126-6708/2003/01/035
[arXiv:hep-ph/0210342 [hep-ph]].

\bibitem{deGouvea:2003xe}
A.~de Gouvea and H.~Murayama,
Phys. Lett. B \textbf{573}, 94-100 (2003)
doi:10.1016/j.physletb.2003.08.045
[arXiv:hep-ph/0301050 [hep-ph]].

\bibitem{Agashe:2008fe}
K.~Agashe, T.~Okui and R.~Sundrum,
Phys. Rev. Lett. \textbf{102}, 101801 (2009)
doi:10.1103/PhysRevLett.102.101801
[arXiv:0810.1277 [hep-ph]].

\bibitem{Jeong:2012zj}
K.~S.~Jeong and F.~Takahashi,
JHEP \textbf{07}, 170 (2012)
doi:10.1007/JHEP07(2012)170
[arXiv:1204.5453 [hep-ph]].

\bibitem{deGouvea:2012ac}
A.~de Gouvea and H.~Murayama,
Phys. Lett. B \textbf{747}, 479-483 (2015)
doi:10.1016/j.physletb.2015.06.028
[arXiv:1204.1249 [hep-ph]].



\bibitem{Altarelli:2012ia}
G.~Altarelli, F.~Feruglio, I.~Masina and L.~Merlo,
JHEP \textbf{11}, 139 (2012)
doi:10.1007/JHEP11(2012)139
[arXiv:1207.0587 [hep-ph]].



\bibitem{Brdar:2015jwo}
V.~Brdar, M.~K\"onig and J.~Kopp,
Phys. Rev. D \textbf{93}, no.9, 093010 (2016)
doi:10.1103/PhysRevD.93.093010
[arXiv:1511.06371 [hep-ph]].



\bibitem{Ge:2018ofp}
S.~F.~Ge, A.~Kusenko and T.~T.~Yanagida,
Phys. Lett. B \textbf{781}, 699-705 (2018)
doi:10.1016/j.physletb.2018.04.040
[arXiv:1803.03888 [hep-ph]].

\bibitem{Barrie:2019pqc}
N.~D.~Barrie, S.~F.~Ge and T.~T.~Yanagida,
Phys. Lett. B \textbf{801}, 135159 (2020)
doi:10.1016/j.physletb.2019.135159
[arXiv:1911.07430 [hep-ph]].



\bibitem{Jeong:2014qpa}
K.~S.~Jeong, N.~Kitajima and F.~Takahashi,
Phys. Rev. D \textbf{91}, no.11, 113010 (2015)
doi:10.1103/PhysRevD.91.113010
[arXiv:1412.4061 [hep-ph]].

\bibitem{Fortin:2016zyf}
J.~F.~Fortin, N.~Giasson and L.~Marleau,
Phys. Rev. D \textbf{94}, no.11, 115004 (2016)
doi:10.1103/PhysRevD.94.115004
[arXiv:1609.08581 [hep-ph]].

\bibitem{Babu:2016aro}
K.~S.~Babu, A.~Khanov and S.~Saad,
Phys. Rev. D \textbf{95}, no.5, 055014 (2017)
doi:10.1103/PhysRevD.95.055014
[arXiv:1612.07787 [hep-ph]].

\bibitem{Fortin:2017iiw}
J.~F.~Fortin, N.~Giasson and L.~Marleau,
JHEP \textbf{04}, 131 (2017)
doi:10.1007/JHEP04(2017)131
[arXiv:1702.07273 [hep-ph]].

\bibitem{Long:2017dru}
A.~J.~Long, M.~Raveri, W.~Hu and S.~Dodelson,
Phys. Rev. D \textbf{97}, no.4, 043510 (2018)
doi:10.1103/PhysRevD.97.043510
[arXiv:1711.08434 [astro-ph.CO]].

\bibitem{Fortin:2018etr}
J.~F.~Fortin, N.~Giasson and L.~Marleau,
Nucl. Phys. B \textbf{930}, 384-398 (2018)
doi:10.1016/j.nuclphysb.2018.03.009
[arXiv:1801.10165 [hep-ph]].


\bibitem{Gell-Mann:1979vob}
M.~Gell-Mann, P.~Ramond and R.~Slansky,
Conf. Proc. C \textbf{790927}, 315-321 (1979)
[arXiv:1306.4669 [hep-th]].

%
\bibitem{Yanagida:1979as}
T. Yanagida, Proceedings: Workshop on the Unified Theories and the Baryon Number in the Universe: Tsukuba, Japan, February 13-14, 1979, Conf. Proc. C7902131, 95 (1979);  Phys. Rev. D 20, 2986 (1979).

\bibitem{Mohapatra:1986aw}
R.~N.~Mohapatra,
Phys. Rev. Lett. \textbf{56}, 561-563 (1986)
doi:10.1103/PhysRevLett.56.561.

\bibitem{Mohapatra:1986bd}
R.~N.~Mohapatra and J.~W.~F.~Valle,
Phys. Rev. D \textbf{34}, 1642 (1986)
doi:10.1103/PhysRevD.34.1642.

\bibitem{Ellis:1992zr}
J.~R.~Ellis, J.~L.~Lopez and D.~V.~Nanopoulos,
Phys. Lett. B \textbf{292}, 189-194 (1992)
doi:10.1016/0370-2693(92)90629-I
[arXiv:hep-ph/9207237 [hep-ph]].


\bibitem{Ellis:1992nq}
J.~R.~Ellis, D.~V.~Nanopoulos and K.~A.~Olive,
Phys. Lett. B \textbf{300}, 121-127 (1993)
doi:10.1016/0370-2693(93)90758-A
[arXiv:hep-ph/9211325 [hep-ph]].

\bibitem{Kang:2006sn}
S.~K.~Kang and C.~S.~Kim,
Phys. Lett. B \textbf{646}, 248-252 (2007)
doi:10.1016/j.physletb.2006.12.071
[arXiv:hep-ph/0607072 [hep-ph]].

\bibitem{Davier:2017zfy} 
  M.~Davier, A.~Hoecker, B.~Malaescu and Z.~Zhang,
  Eur.\ Phys.\ J.\ C {\bf 77}, no. 12, 827 (2017)
  [arXiv:1706.09436 [hep-ph]].
   
\bibitem{Keshavarzi:2018mgv} 
  A.~Keshavarzi, D.~Nomura and T.~Teubner,
  Phys.\ Rev.\ D {\bf 97}, no. 11, 114025 (2018)
  [arXiv:1802.02995 [hep-ph]].

\bibitem{ParticleDataGroup:2020ssz}
P.~A.~Zyla \textit{et al.} [Particle Data Group],
PTEP \textbf{2020}, no.8, 083C01 (2020)
doi:10.1093/ptep/ptaa104


\bibitem{KamLAND-Zen:2016pfg}
A.~Gando \textit{et al.} [KamLAND-Zen],
Phys. Rev. Lett. \textbf{117}, no.8, 082503 (2016)
doi:10.1103/PhysRevLett.117.082503
[arXiv:1605.02889 [hep-ex]].

\bibitem{Agostini:2022zub}
M.~Agostini, G.~Benato, J.~A.~Detwiler, J.~Men\'endez and F.~Vissani,
[arXiv:2202.01787 [hep-ex]].

\bibitem{Akemann:2013}
  G.~Akemann, M.~Kieburg, L.~Wei,
  J. Phys. A: Math. Theor. \textbf{46} 275205(2013) 
[arXiv:1303.5694 [math-ph]].

\bibitem{Gradshteyn} I.~S. Gradshteyn and I.~M. Ryzhik, \textit{Table of Integrals,
Series, and Products}, Academic Press, San Diego  2000.

\end{thebibliography}
\end{document}